\documentclass[10pt, twocolumn]{IEEEtran}
\usepackage[T1]{fontenc}
\usepackage{color}
\usepackage{float}
\usepackage{mathrsfs}
\usepackage{amsmath}
\usepackage{amssymb}
\usepackage{graphicx}
\usepackage{float} 
\usepackage{booktabs} 
\usepackage{caption}  
\usepackage{CJK}
\usepackage{indentfirst}
\usepackage{amsmath}
\usepackage{cases}
\usepackage{setspace}
\usepackage[unicode=true,
bookmarks=true,bookmarksnumbered=true,
bookmarksopen=true,bookmarksopenlevel=1,
breaklinks=false,pdfborder={0 0 0},
pdfborderstyle={},backref=false,
colorlinks=false]
{hyperref}
\hypersetup{pdftitle={Your Title},
pdfauthor={Yulu Han},
pdfpagelayout=OneColumn, 
pdfnewwindow=true, 
pdfstartview=XYZ, 
plainpages=false}
\makeatletter
\floatstyle{ruled}
\newfloat{algorithm}{tbp}{loa}
\providecommand{\algorithmname}{Algorithm}
\floatname{algorithm}{\protect\algorithmname}
\usepackage{setspace}
\usepackage{multirow} 
\usepackage{xcolor}
\usepackage[caption=false,font=footnotesize]{subfig}
\usepackage{algorithm}
\usepackage{algorithmic}
\usepackage{multirow} 
\usepackage{amsmath} 
\usepackage{xcolor}

\usepackage[caption=false,font=footnotesize]{subfig}
\allowdisplaybreaks[4]

\usepackage{cite}
\usepackage{bm}
\usepackage{makecell}
\usepackage{fancyhdr}
\usepackage{diagbox}
\usepackage{algorithmic}
\usepackage{algorithm}
\usepackage{graphicx}
\usepackage{enumitem}
\IEEEoverridecommandlockouts

\usepackage{lettrine}

\usepackage{geometry}
\usepackage{subfig}
\usepackage{hyperref}
\makeatother

\usepackage{hyperref}
\begin{document}

\title{\textcolor{black}{SDN-Blockchain Based Security Routing 
for UAV Communication via Reinforcement Learning}}
\author{
\IEEEauthorblockN{Yulu Han$^{\dagger}$, Ziye Jia$^{\dagger}$, Jingjing Zhao$^{\ddagger *}$, Lijun He$^{\S}$, Yao Wu$^{\dagger}$, and Qihui Wu$^{\dagger}$\\
}
\IEEEauthorblockA{
\small$^{\dagger}$The Key Laboratory of Dynamic Cognitive System of Electromagnetic Spectrum Space, Ministry of Industry and\\ 
\small Information Technology, Nanjing University of Aeronautics and  Astronautics, Nanjing, Jiangsu, 211106, China\\
\small$^{\ddagger}$School of Electronics and Information Engineering, Beihang University, 100191, Beijing, China\\ 
\small$^{*}$The State Key Laboratory of CNS/ATM, 100191, Beijing, China\\
\small$^{\S}$School of Information and Control Engineering, China University of Mining and Technology, Xuzhou, 221116, China\\
\{hanyulu, jiaziye, wu\_yao, wuqihui\}@nuaa.edu.cn, jingjingzhao@buaa.edu.cn, lijunhe@cumt.edu.cn.
}
\thanks{{This work was supported in part by National Natural Science Foundation of China 
under Grant 62301251, in part by the Natural Science Foundation on Frontier Leading 
Technology Basic Research Project of Jiangsu under Grant BK20222001, in part by the 
Open Project Program of State Key Laboratory of CNS/ATM (No.2025B11), and in part 
by the Young Elite Scientists Sponsorship Program by CAST 2023QNRC001. (Corresponding author: Ziye Jia)}}
}
\maketitle
\thispagestyle{empty}
\begin{abstract}
The unmanned aerial vehicle (UAV) network plays important roles in 
emergency communications. 
However, it is challenging to design reliable routing strategies that ensure low latency, energy efficiency, 
and security in the dynamic and attack-prone environments. 
To this end, we design a secure routing architecture integrating 
software-defined networking (SDN) for centralized control and blockchain 
for tamper-proof trust management. In particular, 
a novel security degree metric is introduced to quantify the UAV trustworthiness. 
Based on this architecture, we propose a beam search-proximal policy optimization (BSPPO) algorithm, 
where beam search (BS) pre-screens the high-security candidate paths, and proximal policy optimization 
(PPO) performs hop-by-hop 
routing decisions to support dynamic rerouting upon attack detections. 
Finally, extensive simulations under varying attack densities, packet sizes, and rerouting events 
demonstrate that BSPPO outperforms PPO, BS-Q learning, and BS-actor critic in terms of 
delay, energy consumption, and transmission success rate, showing the outstanding robustness and adaptability.

\end{abstract}

\newcommand{\CLASSINPUTtoptextmargin}{0.8in}

\newcommand{\CLASSINPUTbottomtextmargin}{1in}

\section{Introduction}
\vspace{2mm}
\lettrine[lines=2]THE unmanned aerial vehicles (UAVs) have become a 
vital part of modern wireless communication systems due to their rapid advancement. 
The flexible deployment of UAVs enables 
quick network establishment in disaster scenarios, supporting data collection, real-time 
transmission, and reliable communication for rescue operations.
An effective UAV network architecture ensures reliable communication and 
information sharing, supporting situational awareness and decision-making in 
emergency rescue tasks\cite{HEJIA1}. 
Despite these advantages, the emergency communication in UAV networks may face the 
critical challenges in low latency and energy efficiency while ensuring the 
transmission security under highly dynamic and insecure environments. 

There exist a couple of researches on optimizing routing in UAV networks. 
For instance, traditional shortest-path algorithms such as Dijkstra and Ad-hoc on-demand distance 
vector focus on minimizing the hop count or distance, which perform suboptimally under varying 
network conditions\cite{tr1}. 
To maximize the minimum residual energy of sensors after data transmission in UAV networks, 
\cite{II4} proposes the Voronoi vertex to determine 
hovering location so that the UAV can collect data from more adjacent sensors.
\cite{II3} proposes a novel routing mechanism termed as  
Multi-Objective Markov Decision Based Routing, a Q-learning-based method that enhances the 
network lifetime and packet delivery in search operations. 
At the security level, \cite{II6} proposes a UAV security framework integrating 
dual-channel encryption and authentication to defend against the network and physical hijacking attacks.
However, the existing studies typically address the latency, energy efficiency, or security in isolation, 
lacking a unified approach jointly considering the low delay, low energy consumption, and 
security in dynamic UAV networks. 

To this end, we design a security architecture for UAV networks with adaptive 
routing algorithm, i.e., beam search-proximal policy optimization (BSPPO). 
The designed architecture integrates software-defined networking (SDN) for centralized control and blockchain 
for decentralized and tamper-proof security supervision. 
Specifically, by seperating the control and data planes, SDN improves the management flexibility and 
enables rapid adaptation to network dynamics, facilitating the path adjustments and resource optimization\cite{sdn1,sdn2}.
Besides, due to its decentralization, transparency, and
traceability, the blockchain technology has been widely adopted across various domains as an 
effective solution to enhance both efficiency and security\cite{10453360, hanyulu}. 
The timely security authentication states of UAVs are recorded on the blockchain for the supervision of SDN controllers.
We further introduce the security degree (SD) to quantify the trustworthiness of the UAV node.
Based on the above architecture, we propose a 
BSPPO algorithm to leverage the beam search (BS) to screen the candidate paths and
proximal policy optimization (PPO) to make routing decisions. 
Extensive simulations are conducted in UAV networks
from multiple dimensions including transmission success
rate, delay, and energy consumption to verify the performance, 
which demonstrate the effectiveness and robustness of the proposed architecture and 
algorithm.

The rest of this paper is arranged as follows. 
Section \ref{section2} presents the system model and problem formulation.
In Section \ref{section3}, the security routing algorithm BSPPO is designed. 
We conduct simulations and analyze the results in Section \ref{section4}. Finally, Section \ref{section5} 
draws conclusions.


\begin{figure}[t]
     \centering
     \includegraphics[width=1\linewidth]{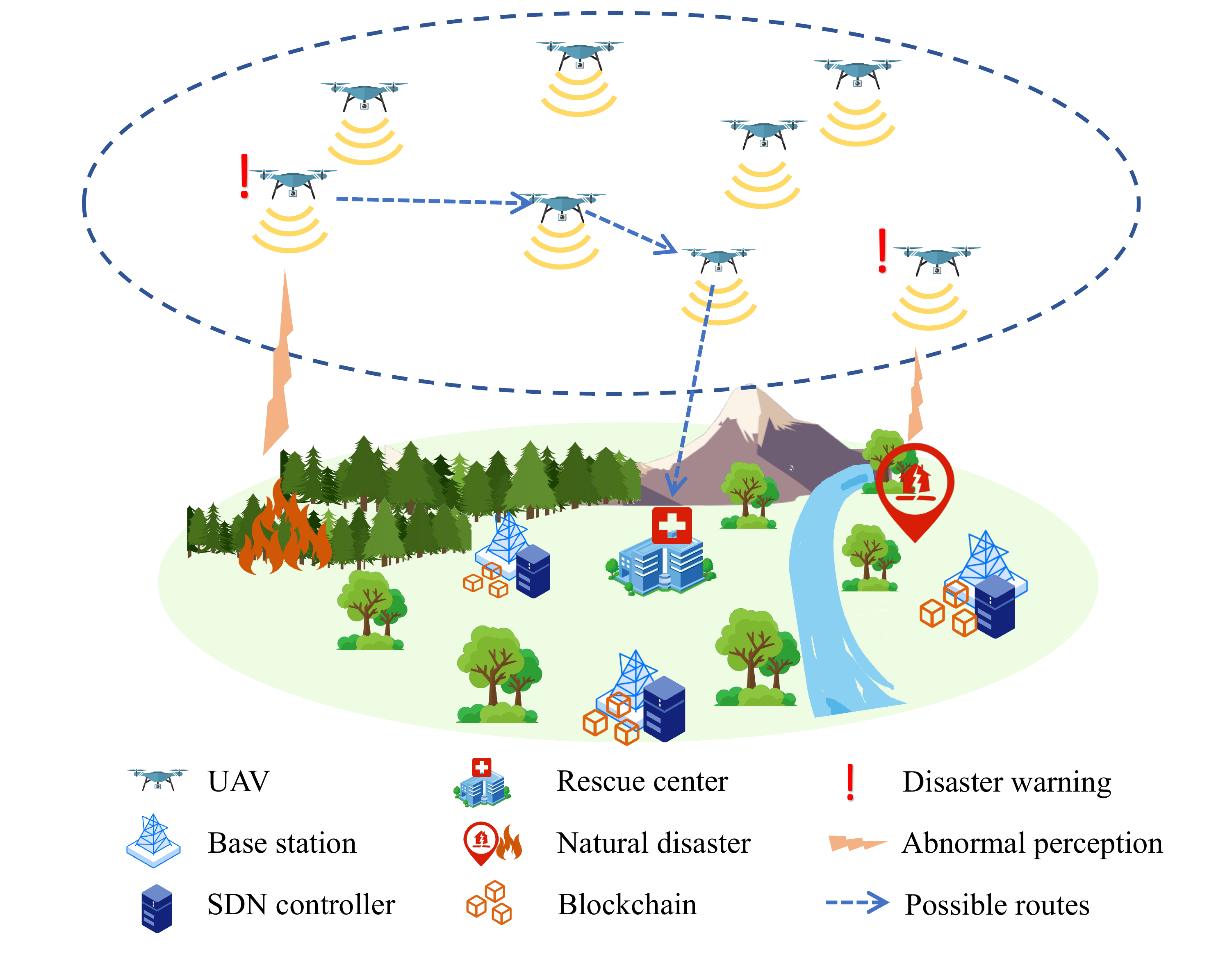}
     \captionsetup{font=small} 
     \caption{\raggedright Routing scenario for emergency rescue.}
     \label{fig:scene}
\end{figure}

\section{System Model and Problem Formulation}\label{section2}
In this section, we present the system model of 
the emergency communication scenario and conduct the problem formulation.

\subsection{Network Model}
As shown in Fig. \ref{fig:scene}, we consider $N$ low-altitude UAVs 
flying over an $M\times M$ area according to the established monitoring trajectory, with 
the index sets of $\mathcal{N}=\{1,2,\ldots,N\}$. 
The UAV network is represented as an undirected graph $\mathcal{G}=(\mathcal{V},\mathcal{E})$, 
where $\mathcal{V}$ represents the set of nodes 
including UAVs $v_i$ and destination rescue center $d$. 
$\mathcal{E}\subseteq\mathcal{V}\times\mathcal{V}$ 
is the set of communication links between nodes, in which $e_{i,j}\in\{0,1\}$ is a binary 
variable. In particular, $e_{i,j}=1$ indicates data transmission from UAVs $i$ to $j$, 
and $e_{i,j}=0$ represents there is no transmission\cite{10051806,10284492}. 
UAVs fly along the established trajectory to monitor the environment with equipped sensors. 
The max transmission range for UAV communication is $O_{\mathrm{max}}$. 
UAVs at each position $p_{\mathrm{pos}}=(x_i(t),y_i(t))$ monitor the area and capture images 
and videos from the surveillance area with constant altitudes. 
In the proposed UAV-assisted emergency communication system, 
UAVs serve as mobile relays to forward environmental monitoring data towards 
the destination rescue center via the SDN-controlled multi-hop network.
When the sensors detect the changes of the environment and possibilities 
of natural disasters such as wildfires, UAVs deliver the detected data to the rescue center.

\vspace{1.2mm}
\subsection{Security Assessment Model based on SDN and Blockchain}
A global security assessment model is constructed by taking advantage of the centralized 
distribution feature of the SDN, where the data plane and the control plane are separated. 
The data related to the disasters and environment implements as the data plane in the 
SDN system when transmitted and relayed by UAVs. 
Meanwhile, UAVs are equipped with lightweight identity security authentication 
modules and connected to the blockchain to broadcast real time security authentication 
results. To maintain the calculation of security assessment and blockchain 
records, three SDN controllers are deployed on the base stations. 
Besides, the trusted routing algorithm deployed on the SDN controllers performs planning and 
dynamic rerouting based on the security assessment results of the SDN controllers. 
The lightweight identity security authentication modules use the 
binary metric $\delta_{\mathrm{data}}\in\{0,1\}$ to determine whether the UAV is attacked. 
$\delta_{\mathrm{data}}$ equaling to 1 indicates that the UAV is attacked, and 0 otherwise. 
To help make routing decisions and assess the security, 
we design the historical credibility $A$, reliability of adjacent UAVs $RE$, 
and final security degree $SD$ to define the security level of each UAV. 
Large $SD$ means the UAV has a low probability of being attacked.

The historical credibility $A_i^{h}$ of UAV $i$ at hop $h$ is defined as 
the ratio of the number of successful authentication attempts to the total number 
of past authentication attempts. 
Let $S_i^{h}$ denote 
the number of successful authentication attempts of UAV $i$, and 
$F_i^{h}$ is the number of authentication failures. 
$\delta_{\mathrm{data}}$ affects the states of $S_i^{h}$ and $F_i^{h}$. 
When $\delta_{\mathrm{data}}=0$, $S_i^{(h+1)}$ is updated as $S_i^{h}+1$, and 
$F_i^{(h+1)}$ is updated as $F_i^{h}+1$ when $\delta_{\mathrm{data}}=1$.
Thus, when the security states change, $A_i^{(h+1)}$ is calculated as 
\begin{equation}
    A_i^{(h+1)}=\frac{S_i^{(h+1)}}{S_i^{(h+1)}+F_i^{(h+1)}}.
\end{equation}

We define the set of adjacent UAVs as $\boldsymbol{\Gamma}_i$ and
the reliability of every adjacent UAV as $RE_{j}$, where $j\in\boldsymbol{\Gamma}_i$.
A higher $RE$ value means the adjacent UAV is more reliable. 
Hence, the average reliability of adjacent $RE_i$ of 
UAV $i$ is calculated as 
\begin{equation}
    RE_i=\frac{1}{|\boldsymbol{\Gamma}_i|}\sum_{j\in\Gamma(i)}RE_j^{h}.
\end{equation}
$|\boldsymbol{\Gamma}_i|$ represents the number of the adjacent UAVs $i$ 
and $RE_j^{h}$ denotes the reliability of UAV $j$ at hop $h$\cite{3e1}. 
$RE_{j}$ ranges from 0 to 1. 
The initialization of $RE_j$ is informed by historical observations. 
Similarly, the reliability of adjacent UAV changes along with $\delta_{\mathrm{data}}$, i.e.,
\begin{equation}
    RE_j^{(h+1)}=RE_j^{h}-\beta\cdot\delta_{\mathrm{data}}.
\end{equation}
$\beta$ works as the rate of decline to avoid excessive penalties.

The security degree $SD_i$ of UAV $i$ is formulated as a weighted combination of 
the reliability of adjacent UAVs and its historical credibility, integrating both factors as 
\begin{equation}
    SD_i=\alpha RE_i+(1-\alpha) A_i,
\end{equation}
where $\alpha= 0.5$ can balance the influence of 
$RE_i$ and $A_i$ on the SD. 

\vspace{1.2mm}
\subsection{Communication Model}    
We utilize the free-space propagation model for the UAVs communication. 
The free space path loss between UAVs is detailed as  
\begin{equation}
    \begin{aligned}
        FSPL=20log_{10}(\frac{4\pi d_{ij}(t)f}c),
    \end{aligned}
\end{equation}
where $f$ denotes the frequency and $c$ is the speed of light\cite{3b2}. The 
distance $d_{ij}(t)$ between UAVs is represented as 
$d_{ij}(t)=\sqrt{\left(x_{i}(t)-x_{j}(t)\right)^{2}+\left(y_{i}(t)-y_{j}(t)\right)^{2}}$.
$PL_{ij}$ in $dB$ represents path loss using free-space propagation model from the UAVs $i$ 
to $j$, which is calculated as 
\begin{equation}
    \begin{aligned}
        PL_{i,j}=20log_{10}(d_{ij}(t))+20log_{10}(f)-147.55.
    \end{aligned}
\end{equation}
According to the Shannon theorem, the max transmission rate is expressed as 
\begin{equation}
    \mathcal{R}_{i,j}=B_{i,j}log_2(1+SNR_{i,j}),
\end{equation}
where $B_{i,j}$ is the channel bandwidth, and $SNR_{i,j}$ is the signal-to-noise ratio (SNR) in the 
communication channel between the UAV $i$ and $j$, i.e.,
\begin{equation}
    SNR_{i,j}=\frac{\mathcal{P}_{i,j}10^{-\frac{PL_{i,j}}{10}}}{\sigma_{i,j}^2},
\end{equation}
where $\sigma_{i,j}^2$ denotes the noise power between the UAVs $i$ and $j$. 
$\mathcal{P}_{i,j}$ represents the transmission power.

\vspace{1.5mm}
\subsection{Energy Consumption Model in UAV Network}
We focus on optimizing the communication energy consumed 
during UAV routing, as flight and monitoring energy remain relatively constant.
We model the primary energy cost as hop-by-hop packet forwarding, 
including transmission energy $E^{T}$ and reception energy $E^{R}$ 
under a free-space wireless channel model\cite{3c1}.
For the data sent from UAV $i$ to UAV $j$, the per-hop energy is
\begin{equation}
    E_{ij}^{T}=L_{\mathrm{data}} E_e+\mathcal{P}_{i,j} \frac{L_{\mathrm{data}}}{B_{i,j}},
\end{equation}
and 
\begin{equation}
    E_{ij}^{R}=L_{\mathrm{data}} E_e,
\end{equation}
where $L_{\mathrm{data}}$ in $bits$ is the data size, and $E_e$ is the per-bit circuitry energy.
Then, the total routing energy cost $E_{\mathrm{tot}}$ is 
\begin{equation}
    E_{\mathrm{tot}}=\sum_{i,j\in U_{\mathrm{p}}}E_{ij}^{T}+E_{ij}^{R},
\end{equation}
where $U_{\mathrm{p}}$ denotes the candidate path set which contains all the UAVs in the 
route for emergency transmission.
\vspace{1.5mm}
\subsection{Delay Model}

The delay analysis is essential for guaranteeing the timeliness and reliability of the 
mission in the UAV emergency communication scenario. The total delay in routing 
process consists of five parts: 
the request delay, process delay, response delay, end-to-end delay, and the final destination 
delay. 
The request delay $T_{\mathrm{req}}$ represents the propagation delay for the source UAV 
to detect an event and upload the message on the blockchain. 
The process delay $T_{\mathrm{proc}}$ is given as 
\begin{equation}
    T_{\mathrm{proc}}=t_{\mathrm{quer}}+T_{\mathrm{sec}}, 
\end{equation}
where $t_{\mathrm{quer}}$ and $T_{\mathrm{sec}}$ represent the query time and processing time 
of the SDN, respectively. The SDN controllers query the consensus information such as 
the IDs and locations of UAVs on the blockchain. Let $t_{\mathrm{comp}}$ 
represent the time for the SDN to calculate the SD for each UAV and generate the response delay. 
Then, the total processing time is calculated as
\begin{equation}
    T_{\mathrm{sec}}=\sum_{n=1}^Nt_{\mathrm{comp}}=\sum_{n=1}^{N}\frac{C_n}{f_{\mathrm{SDN}}},
\end{equation}
where $C_n$ indicates the computing load of each UAV, and $f_{\mathrm{SDN}}$ represents the clock 
frequency of the SDN controller\cite{10735222}. 
Furthermore, the SDN controller transmits the calculated routing information back to the source UAV,  
which generates the response delay $T_{\mathrm{resp}}$, i.e., 
\begin{equation}
    T_{\mathrm{resp}}=\frac{L_{\mathrm{path}}}{B_{i,S_j}},
\end{equation}
where $L_{\mathrm{path}}$ denotes the size of routing packets and $B_{i,S_j}$ is the bandwidth between 
the source UAV $i$ and SDN controller $j$.
The end-to-end delay between UAVs includes the transmission, and self-authentication. 
Specifically, the routing delay between UAVs $i$ and $j$ is expressed as 
\begin{equation}
    T_{i,j}=t_{\mathrm{proc},i}+\frac{L_{\mathrm{data}}}{\mathcal{R}_{i,j}}+\frac{d_{i,j}}{c},
    \label{eq:tij}
\end{equation}
where $t_{\mathrm{proc,i}}$ represents the time for self-authentication on UAV $i$.
Thus, the total end-to-end delay can be calculated by
\begin{equation}
    T_{\mathrm{eted}}=\sum_{i,j\in U_{\mathrm{p}}}e_{i,j}T_{i,j}.
\end{equation}
The queuing delay is assumed to be ignored since 
the emergency communication is usually treated as a separate transmission primarily.
Let $T_{\mathrm{des}}$ represent the delay from the last UAV to the destination rescue center, 
which is calculated as 
\begin{equation}
    T_{\mathrm{des}}=\frac{L_{\mathrm{data}}}{\mathcal{R}_{i,j}}+\frac{d_{i,j}}{c}.
\end{equation}
The total delay of the safe routing process can be expressed as
\begin{equation}
    \mathcal{T}_{\mathrm{original}}=T_{\mathrm{req}}+T_{\mathrm{proc}}+T_{\mathrm{resp}}+T_{\mathrm{eted}}+T_{\mathrm{des}}.
\end{equation}

However, once the original route is deemed unsafe, a reroute process is triggered to avoid the attacked UAVs.
The SDN recalculate the new routing path and broadcast it to the previous 
safe UAVs for retransmission with the data cache according to the real-time supervision on the blockchain.
After verification, each UAV node updates the ledger in the blockchain.
The rerouting delay includes three parts as\\
\begin{equation}
    T_{\textnormal{re-routing}}^{(n)}=T_{\mathrm{proc}}^{(n)}+T_{\mathrm{resp}}^{(n)}+T_{\mathrm{eted}}^{(n)}.
\end{equation}
Specifically, $n\in \{0, 1, 2, \dots, k\}$, which denotes the $n$-th detected attack. 
$k$ is the number of detected attacks. 
The rerouting delay with $k$ detected attacks is  
\begin{equation}
    \mathcal{T}_{\textnormal{re-routing}}=T_{\textnormal{re-routing}}^{(1)}+T_{\textnormal{re-routing}}^{(2)}+\ldots+T_{\textnormal{re-routing}}^{(k)}.
\end{equation}
Hence, the total delay of the UAV rescue communication with detected attacks 
can be calculated as
\begin{equation}
    \mathcal{T}=
    \begin{cases}
        \mathcal{T}_{\mathrm{original}},&\mathrm{if}\, \delta_{\mathrm{data}}=0\\
        T_{\mathrm{req}}+T_{\mathrm{proc}}+T_{\mathrm{resp}}+t_{\mathrm{eted}}+T_{\mathrm{des}}\\
        +\delta_{\mathrm{data}}\mathcal{T}_{\textnormal{re-routing}},&\mathrm{if}\, \delta_{\mathrm{data}}=1
    \end{cases},
\end{equation}
where $t_{\mathrm{eted}}$ represents the end-to-end delay before interruptions.

\subsection{Problem Formulation}
To achieve timely and energy-efficient data delivery with dynamic topology and 
security threats, we formulate the routing objective problem that minimizes 
the weighted sum of communication delay and energy consumption, ensuring efficient 
and sustainable emergency response, detailed as follows.
\begin{subequations} \label{P0}
    \begin{align}
    \mathbf{P0}:\quad &\min_{p_b} \quad \lambda \mathcal{T} + (1-\lambda) E_{\text{tot}} \label{P0a} \\
    \text{s.t.} \quad 
    & SNR_{i,j} \geq SNR_{\text{min}}, \quad \forall (i,j) \in \mathcal{N}, \label{P0b} \\
    & SD_i \geq \theta_{\text{SD}}, \quad \forall i \in U_{\mathrm{p}}, \label{P0c} \\
    & E_{\text{residual},i} \geq E_{\text{threshold}}, \quad \forall i \in U_{\mathrm{p}}, \label{P0d} \\
    & d_{ij} \leq O_{\mathrm{max}}, \quad \forall (i,j) \in \mathcal{N},  \label{P0e}\\
    & p_b\in U_{\mathrm{p}}. \label{P0f}
\end{align}
\end{subequations}
Constraint \eqref{P0b} ensures the stable transmission, which indicates that the $SNR_{i,j}$ of each 
selected communication link cannot less than a minimum threshold $SNR_{\mathrm{min}}$. 
Moreover, in \eqref{P0c}, only UAVs with a $SD_i$ value higher than a predefined threshold $\theta_{\text{SD}}$
are allowed to act as relay nodes. Additionally, constraint \eqref{P0d} guarantees that 
relay nodes must maintain a minimum residual 
energy level to prevent premature node failures. \eqref{P0e} guarantees that all the 
transmission distances are within the communication range of UAVs. 
Specifically, $U_{\mathrm{p}}=\{p_1, ..., p_b\}$, and $b$ denotes 
the $b$-th available path. 
The coupling of delay, energy, and security constraints makes 
the routing problem highly complex under dynamic network conditions. 
To efficiently obtain adaptive and secure routing decisions, 
we propose the BSPPO algorithm that combines heuristic path screening with reinforcement learning.

\section{Algorithm Design}\label{section3}
\vspace{2mm}
We propose a two-sequential algorithm termed as BSPPO, as shown 
in the Algorithm \ref{algorithm-BSPPO}. 
We use the BS to preselect high-security UAV and communication 
links, and PPO make adaptive routing 
decisions to ensure both security and efficiency in the UAV-based emergency rescue system.
The cost of a selected path $\mathbb{P}=(v_1,\ldots,v_k,d)$ is expressed as
\begin{equation}
    \mathrm{Score}_\mathrm{(avg)}(\mathbb{P}) = \frac{1}{|\mathbb{P}|} \sum_{\substack{v_i \in \mathbb{P}}} SD_i .\label{score}
\end{equation}
To balance the security and path diversity, we utilize the average security score to retain more 
feasible nodes and edges while ensuring the path security. 
The directional screening restrictions towards destination is applied 
to improve the screening efficiency in the BS. 
Starting from the source node, BS iteratively expands partial paths that satisfy $\mathbf{P0}$ constraints, 
scores them based on \eqref{score}, and keeps the top-$B$ candidates at each hop until reaching the destination. 
In the end, the final beam set $\mathcal{B}_{H}$ which contains $K$ high quality candidate 
paths with the max 
$H$ hops is formed. The set of all selected path is represented as $\mathcal{P}_{\mathrm{B}}$.
We merge all the nodes and edges from the candidate $\mathcal{P}_{\mathrm{B}}$ to obtain 
$\mathcal{G}^{\prime}=(V_{\mathrm{opt}},E_{\mathrm{opt}})$, 
which serves as the action space input of the PPO. 
$V_{\mathrm{opt}}$ and $E_{\mathrm{opt}}$ 
represent the set of nodes and communication links screened by BS, respectively.

\begin{algorithm}[t!]
\caption{\label{algorithm-BSPPO}{BSPPO-based Secure UAV Routing}}
\begin{algorithmic}[1]
\REQUIRE Graph $\mathcal{G}$, source $s$, destination $d$, beam width $B$, max hops $H_{\max}$, episodes $N_{\mathrm{ep}}$, learning rate $\eta$, and clip $\epsilon$.
\STATE \textbf{BS Path Screening:} Initialize $\mathcal{B}_0 \!=\! [[s]]$, $\mathcal{P}_{\mathrm{B}}\!=\!\emptyset$. 
\FOR{$h=1:H_{\max}$}
  \STATE Generate candidate set $\mathcal{C}$ .
  \STATE Add $p\!\in\!\mathcal{B}_h$ ending at $d$ into $\mathcal{P}_{\mathrm{B}}$.
\ENDFOR
\STATE Construct subgraph $\mathcal{G}'$ from $\mathcal{P}_{\mathrm{B}}$.
\STATE \textbf{PPO Training:} Initialize actor $\pi_\theta$, critic $V_\phi$.
\FOR{episode $=1:N_{\mathrm{ep}}$}
  \STATE Reset env on $\mathcal{G}'$, clear buffer.
  \FOR{hop $=1:H_{\max}$}
    \STATE Observe $s_h$, sample $a_h\!\sim\!\pi_\theta(\cdot|s_h)$ from valid actions, execute, and get $r_h$ and $s_{h+1}$.
    \IF{attack is detected}
      \STATE Update security state, reroute, rollback if $\mathcal{G}'$ lacks candidate paths; terminate if reroute count exceeds the limitation.
    \ENDIF
    \STATE Store $(s_h,a_h,r_h,s_{h+1})$.
  \ENDFOR
  \STATE Compute advantages $A_t$.
\ENDFOR
\ENSURE Learned policy $\pi_\theta$ for secure routing.
\end{algorithmic}
\end{algorithm}

\vspace{-0.0cm}
\color{black}

Then, based on $\mathcal{G}^{\prime}$ selected by BS, 
we strive to solve the problem via an efficient reinforcement learning method PPO. 
The per-hop UAV routing is modeled as a Markov Decision Process with defined 
state, action spaces, policy, and reward function. At each step, the agent observes 
its local neighborhood and selects the next-hop relay UAV to jointly minimize transmission 
delay and energy consumption\cite{11122503}. 

\subsubsection{State Space}The topological positions of UAVs are assumed unchanged due to 
the short time of discovery and transmission when UAVs follow relatively fixed  
trajectories and fly at a slow speed. 
Therefore, the state at the time step $t$ includes the location $p_{\mathrm{pos}}^i$, legal neighbor nodes 
$v^i_{\mathrm{leg_j}}$, $SD_i$ and the cost of possible links $c^i_{\mathrm{link_j}}$.
\subsubsection{Action Space}At step $t$, the agent from UAV $i$ selects its next-hop UAV $j$ from the feasible 
candidate set, so the action space is defined as 
\begin{equation}
    a_t\in \{v^i_{\mathrm{leg_j}}, i,j \in V_\mathrm{opt}\}.
\end{equation}

\subsubsection{Reward Function}The reward function is aligned with the objective function in 
$\mathbf{P0}$. Thus, the total reward is designed as 
\begin{equation}
    R_t=-\sum_{k=0}^{T-t}e^k_{i,j}(\lambda^r \tilde{T}_{i,j} + (1-\lambda^r) \tilde{E}_{i,j}),
\end{equation}
where $\tilde{T}_{i,j}$ and $\tilde{E}_{i,j}$ represent the normalized transmission 
delay and energy consumption between nodes $i$ and $j$, respectively. $t\in\{0,...,T\}$ is the 
time step in the training process of PPO.

\subsubsection{Strategy Network and Objective Function}
The policy network $\pi_\theta(a_t|s_t)$ denotes the input state, which outputs the action probability 
distribution, while the value function network $V_\phi(s_t)$ estimates the expected 
total return $R_t$. The objective function is 
\begin{equation}
    L^{\mathrm{CLIP}}(\theta)=\mathbb{E}_t\left[\min\left(r_t(\theta)\hat{A}_t,\mathrm{clip}(r_t(\theta),1-\epsilon,1+\epsilon)\hat{A}_t\right)\right],
\end{equation}
where $r_t(\theta)=\frac{\pi_\theta(a_t|s_t)}{\pi_{\theta_\mathrm{old}}(a_t|s_t)}$ 
is the probability ratio between the new policy $\pi_\theta$ and old 
policy $\pi_{\theta_\mathrm{old}}$. $\hat{A}_t$ is the advantage estimate at time 
step $t$, and 
$\epsilon$ constrains the policy deviation, with clipping preventing overly large 
updates by keeping $r_{t}(\theta)$ close to 1.

\section{Simulation and Analysis}\label{section4}
\vspace{2mm}
We construct the simulation environment using Python 3.10 and Gymnasium, 
and train PPO from Stable-Baselines3 to evaluate the proposed secure routing algorithm.
We consider three baselines including PPO, BS-Q learning (BSQL), 
and BS-actor critic (BSA2C). 
The environment contains 40 UAVs in an $1,200 \times 1,200$ m$^2$ area, where the max communication 
radius $O_\mathrm{max}$ of UAVs is 200 m. 
The SD parameters are set to $\alpha=0.5$, $\beta=0.2$, and $\lambda=0.5$ 
to balance the effects of delay and energy. 
The BS beam width is 60. 
PPO is trained with the learning rate of $5\times10^{-4}$, 
gradient clipping rates of 0.15, $2,048$ time steps, batch size of $64$, and $\epsilon=0.1$.

\begin{figure}[t]
     \centering
     \includegraphics[width=0.78\linewidth]{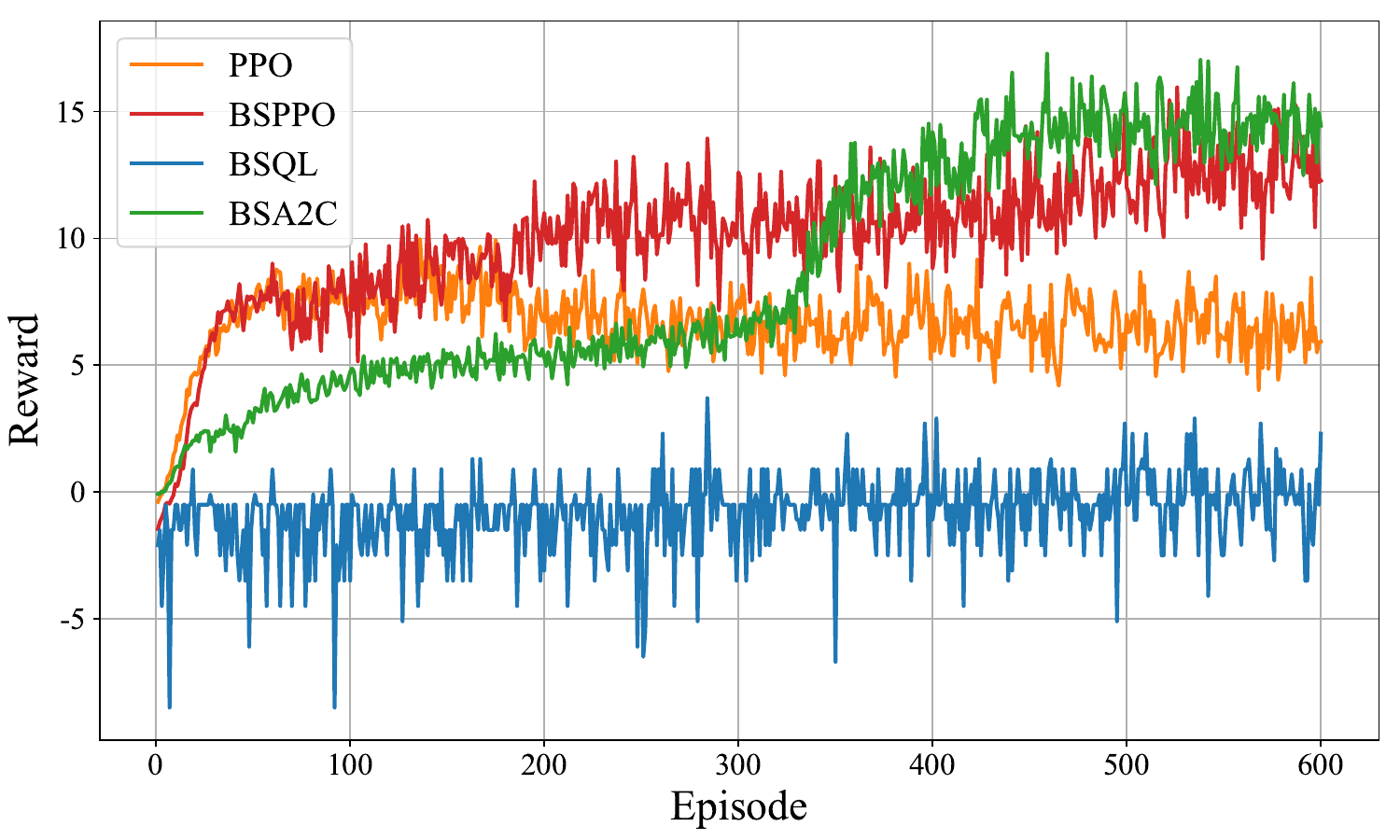}
     \captionsetup{font=small} 
     \caption{\raggedright Rewards and convergence performance under the different algorithms.}
     \label{fig:train_rewards}
\end{figure}

Fig. \ref{fig:train_rewards} shows the convergence of each algorithm with certain number
of attack nodes in the environment. 
It is expected that introducing random attack nodes increases the environmental randomness, 
causing the fluctuations of reward curves. As observed, PPO converges slower than BSPPO for 
the lack of the BS. 
The higher rewards after convergence of BSA2C indicates that it requires more attempts 
to reach the destination, showing lower efficiency than BSPPO.
However, the 
value-based BSQL fails to converge within 600 episodes under such high-dimensional 
randomness and typically needs over 10,000 episodes. 
Therefore, only the first 600 are selected and shown in the figures for comparisons.
\begin{figure}[t]
    \centering
    \subfloat[Delay performance of different algorithms under attack and safe scenarios.]{%
        \includegraphics[width=0.74\linewidth]{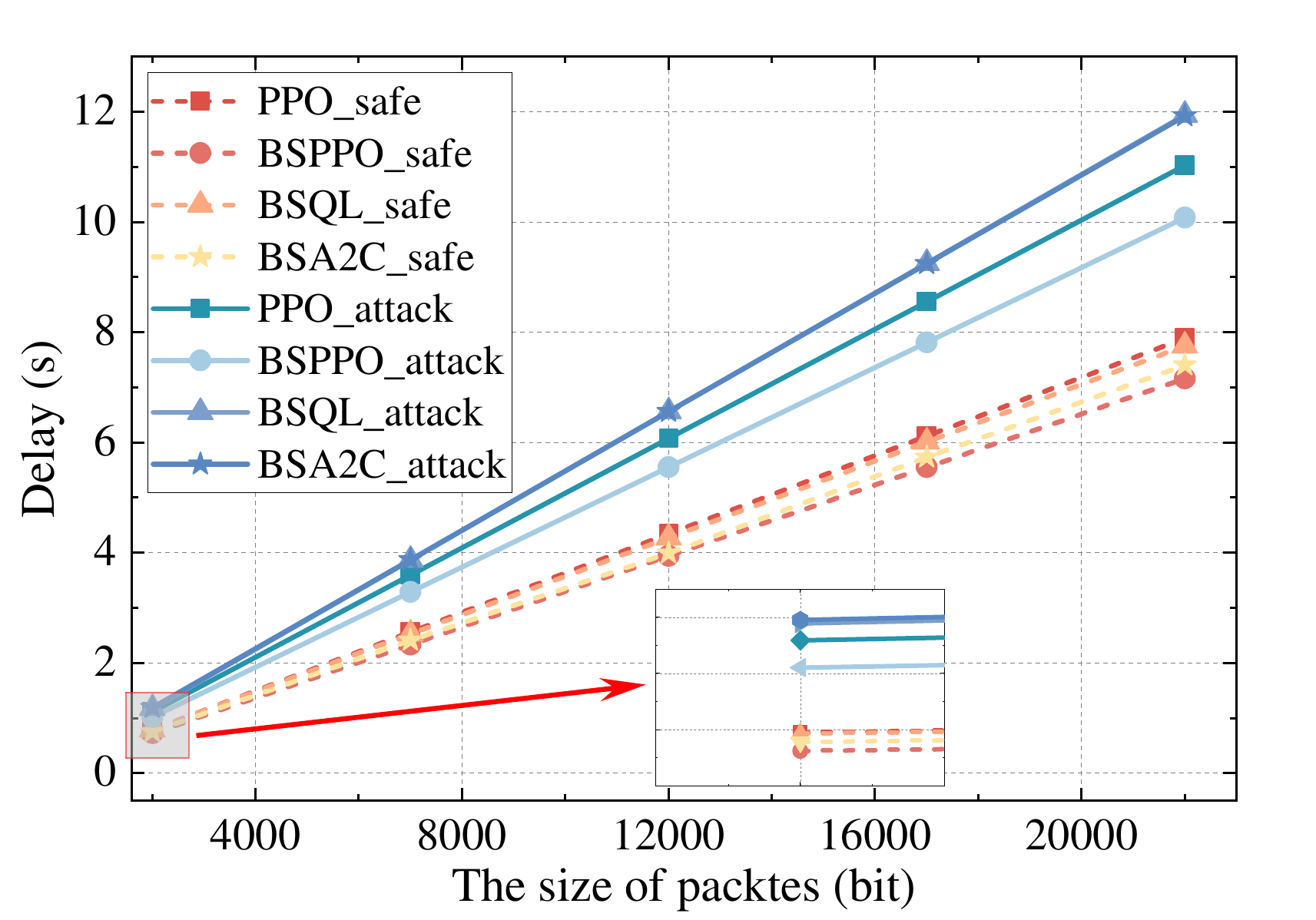}
        \label{fig:delay_attack}
        }\\
    \subfloat[Energy consumption performance of different algorithms under attack and safe scenarios.]{%
        \includegraphics[width=0.74\linewidth]{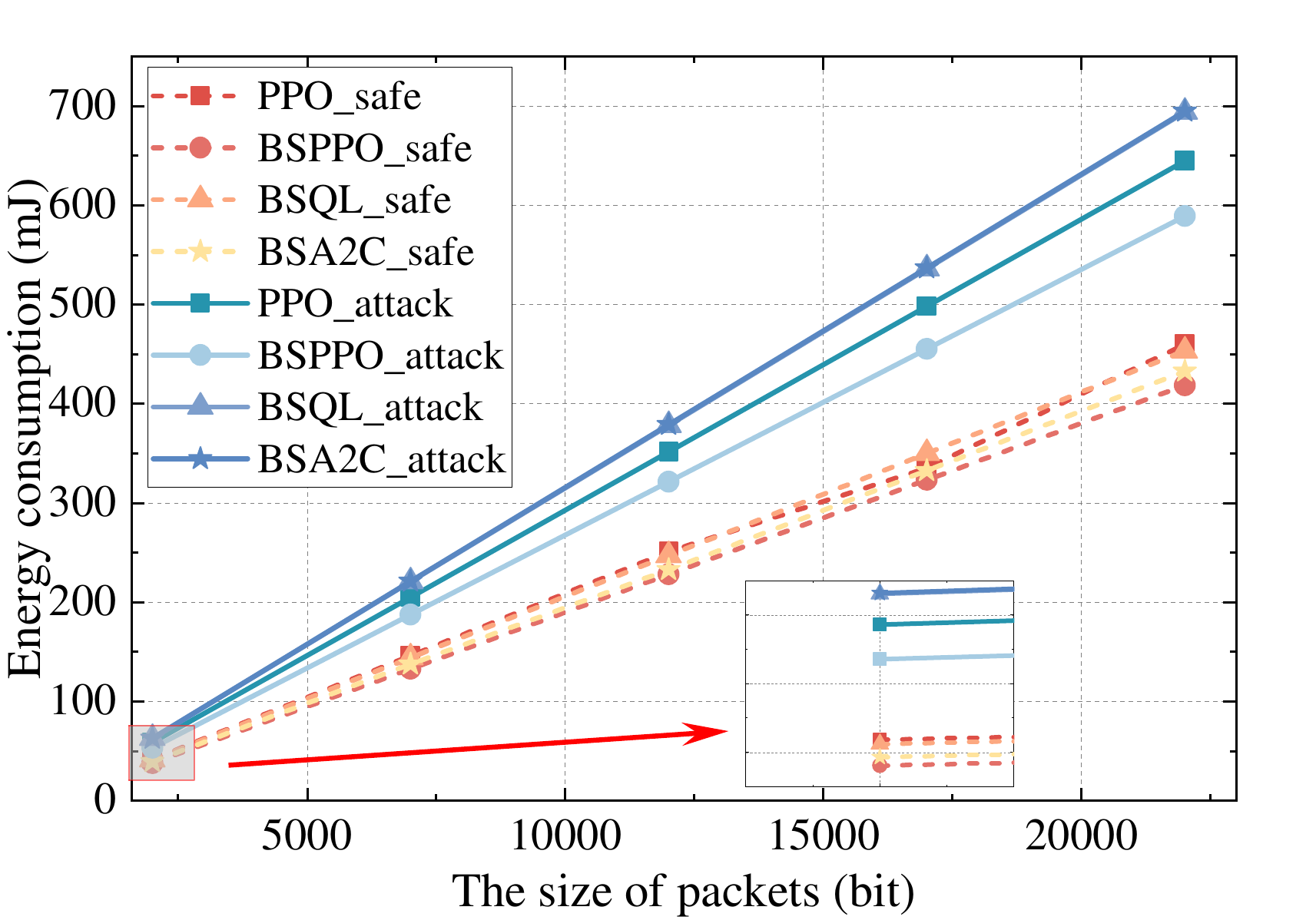}
        \label{fig:energy_attack}
        }
    \caption{Performances of different algorithms with or without attacks.}
    \label{fig:with_attack_comparison}
\end{figure}
\begin{figure}[t]
     \centering
     \includegraphics[width=0.85\linewidth]{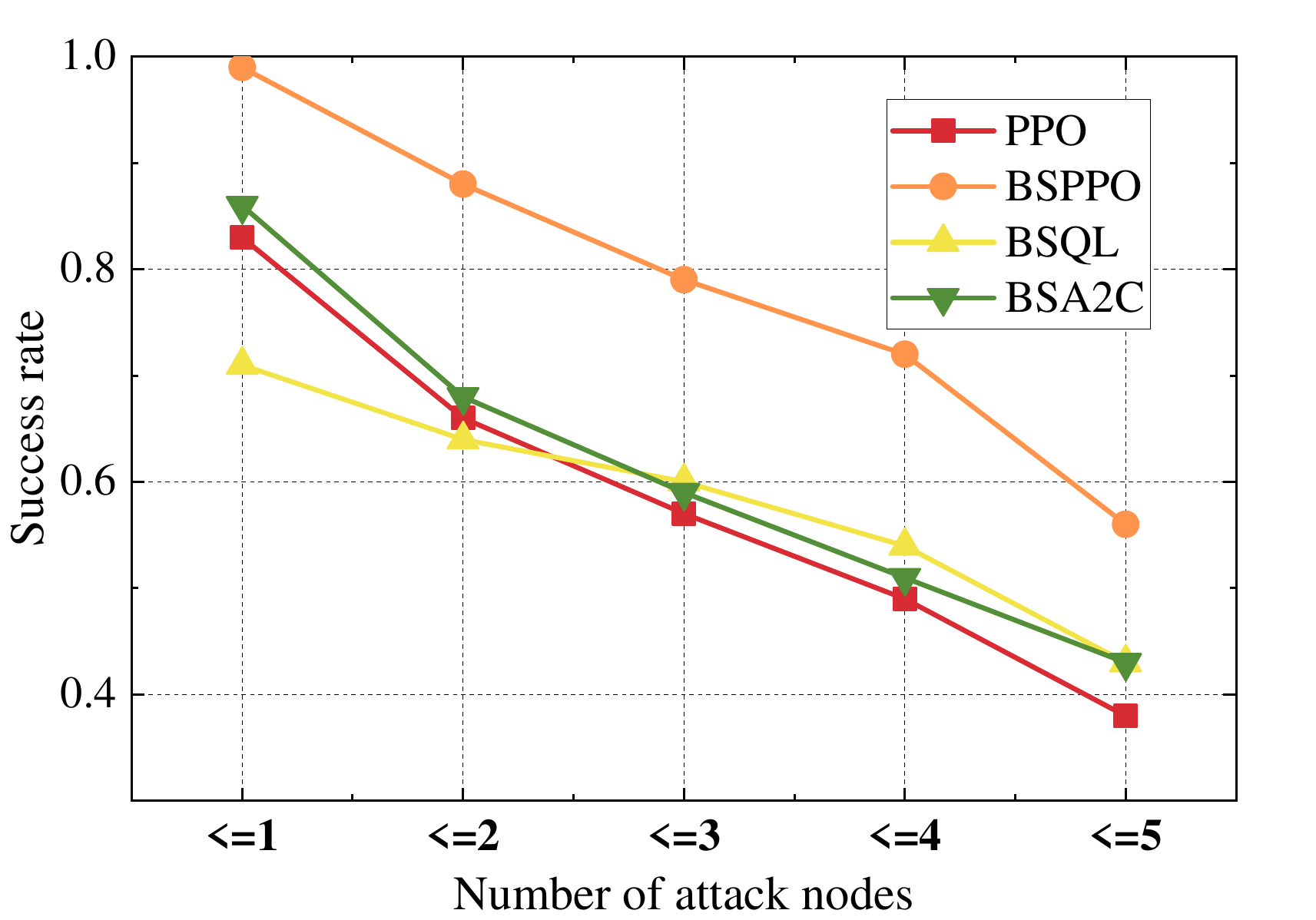}
     \captionsetup{font=small} 
     \caption{\raggedright Transmission success rate of different algorithms under the varying numbers of attack nodes.}
     \label{fig:success_rate_numattacks}
\end{figure}

\begin{figure}
    \centering
    \subfloat[]{%
        \includegraphics[width=0.48\linewidth]{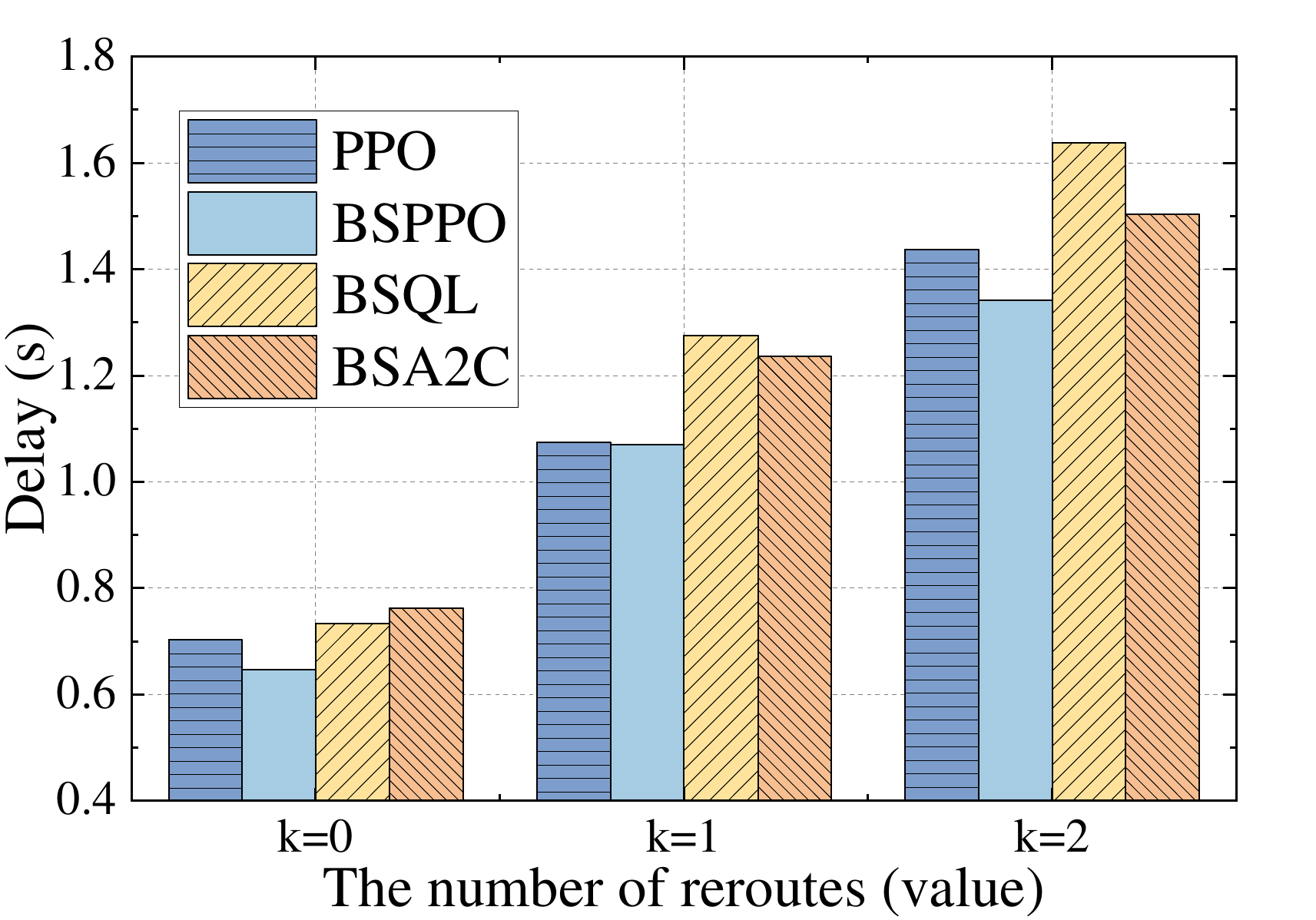}
        \label{fig:delay_attack_k}
        }
    \hfill
    \subfloat[]{%
        \includegraphics[width=0.48\linewidth]{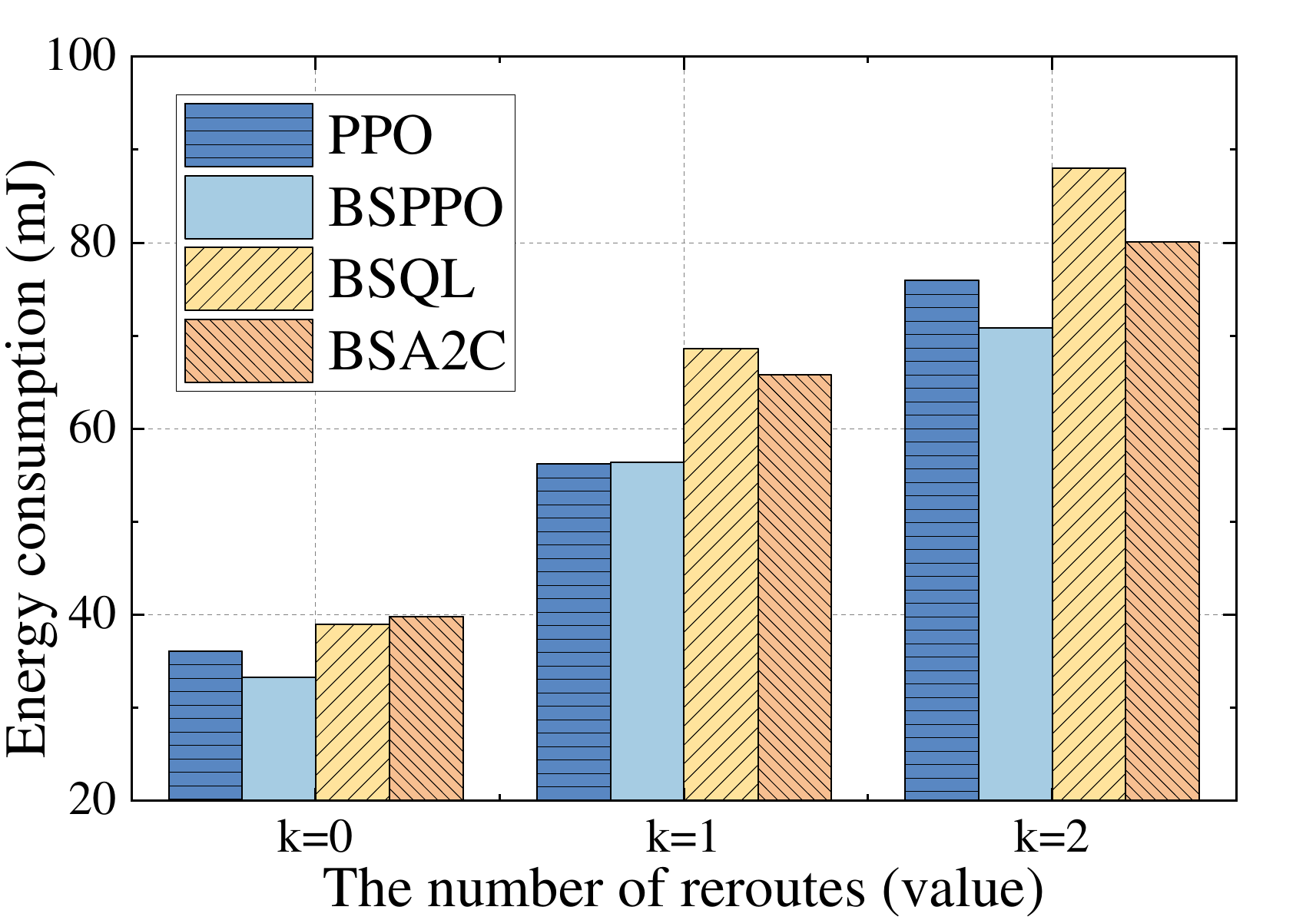}
        \label{fig:energy_attack_k}
        }
    \caption{Delay and energy consumption performances with attacks of different algorithms under different number of reroutes.}
    \label{fig:reroutes_comparison} 
\end{figure}

The comparative results on delay and energy consumption are illustrated in 
Fig. \ref{fig:with_attack_comparison}. 
We evaluate the BSPPO under two scenarios of with attacks and without attacks by varying 
the data packets. 
The environment without attacks is deemed to be safe, which has lower average delay and energy 
consumption as expected since the routing process proceeds smoothly without the 
need for rerouting. 
Conversely, the algorithms must perform rerouting and retransmitting when attacked, which inevitably 
cost more time and energy. 
In both cases, the proposed BSPPO algorithm 
consistently outperforms the baseline algorithms, demonstrating superior efficiency 
and robustness in the securing reliable data delivery.

The secure UAV routing requires dynamic adaptation to the changing attack nodes while 
ensuring reliable data delivery, making the transmission success rate a key performance indicator.
As shown in Fig. \ref{fig:success_rate_numattacks}, 
the transmission success rate of each algorithm declines when the number of attack nodes in 
the environment increases. Nevertheless, the proposed BSPPO consistently achieves the 
highest success rate under all attack intensities, highlighting its superior 
adaptability and robustness in hostile environments.

To study the performances of different algorithms in terms of delay and energy consumption, 
we further evaluate their behaviors under varying numbers of rerouting times, 
which correspond to the number of detected attack nodes along the transmission path. 
Each rerouting removes the previously identified attack nodes from the topology connection graph. 
Excessive rerouting may affect the connectivity of network topology due to the limited sizes 
of the network topology, resulting in the failure of data transmission. 
Therefore, we only study the situation of 0, 1, and 2 times of rerouting. 
As illustrated in Fig. \ref{fig:reroutes_comparison}, 
the proposed BSPPO algorithm consistently exhibits strong performance on cases with 0, 
1, and 2 rerouting times, especially when the reroute times equals to 2, with the lowest delay and energy consumption.

These results collectively demonstrate that BSPPO strikes a favorable balance among the security, 
timeliness, energy efficiency, and robustness, making it well-suited for UAV-assisted emergency 
communication networks operating in dynamic and hostile environments.
\vspace{2mm}

\section{Conclusions}\label{section5} 
In this paper, we propose a secure UAV emergency communication architecture that 
integrates SDN and blockchain with a BSPPO-based routing algorithm. 
By introducing the dynamic SD, the architecture combines the BS to pre-screen path 
and PPO to make hop-by-hop decisions to achieve low latency, energy consumption, 
and high security under dynamic topologies and cyberattacks. 
Extensive simulations under different attack intensities and packet sizes show that 
BSPPO consistently outperforms PPO, BSQL, and BSA2C in terms of the delay, energy efficiency, 
and transmission success rate, demonstrating its robustness and adaptability.

\vspace{-2mm}

\bibliographystyle{IEEEtran}
\bibliography{REFERENCES.bib}
\end{document}